\documentclass{ws-p8-50x6-00}
\usepackage{fleqn,epsfig,amssymb,amsbsy,xspace}

     \newcommand{\pathnow}{}

\newcommand{\myfig}[7]{%
\begin{figure}[#7]
\vskip #5cm	\centerline{\hspace*{#1cm}
\epsfig{width=#2cm,angle=#4,figure=\pathnow #3.ps}
	}\vskip #6cm
                    }
\newcommand{\capt}[3]%
{\caption[{#2}]{\label{Fig:#1}#3}
}
\newcommand{\rf}[1]{Fig.\,\ref{Fig:#1}%
}
\def\qgp{quark--gluon plasma\xspace}

\newcommand{\req}[1]{Eq.\,(\ref{eq:#1})%
}

\newcommand{\beql}[1]{
	\begin{equation} \label{eq:#1}}

\newcommand{\beqarl}[1]{
	\begin{eqnarray} \label{eq:#1} }
\newcommand{\eeql}[1]{\label{eq:#1} \end{equation} 
} 
\newcommand{\eeqarl}[1]{\label{eq:#1} \end{eqnarray} 
}
\newcommand{\rt}[1]{table~\ref{Tab:#1}%
}
\def\beq{\begin{equation}}
\def\eeq{\end{equation}}

\def\beqar{\begin{eqnarray}}
\def\eeqar{\end{eqnarray}}
\def\bcite{\cite}
\def\agev{{$A$~GeV}\xspace}

\newcommand{\captt}[3]%
{\caption[{#2}]{\label{Tab:#1}#3%
}%
}
\newcommand{\lsec}[1]{\label{sec:#1} } 
\newcommand{\res}[1]{section~\ref{sec:#1}%
}

\newcommand{\lssec}[1]{\label{ssec:#1} 
 }

\def\ie{{i.e.\xspace}}
\def\eg{{e.g.\xspace}}

\newcommand{\myfigd}[9]{%
\begin{figure}[#9]
\vskip #5cm	\centerline{\hspace*{#1cm}
\psfig{width=#2cm,angle=#4,figure=\pathnow #3.ps}
\hspace*{#7cm}
\psfig{width=#2cm,angle=#4,figure=\pathnow #8.ps}
	}\vskip #6cm
                    }
\title{PROBING DENSE MATTER WITH STRANGE HADRONS}

\author{Johann Rafelski}

\address{Department of Physics,
University of Arizona, Tucson, AZ 85721\\
and\\ CERN-Theory Division, 1211 Geneva 23, Switzerland}
\author{Jean Letessier}

\address{
Laboratoire de Physique Th\'eorique et Hautes Energies
\\
Universit\'e Paris 7, 2 place Jussieu, F--75251 Cedex 05.
}
\begin{document}


\maketitle

\abstracts{
Analysis of 
hadron production experimental data allows to understand the 
properties of the dense matter fireball 
produced in relativistic heavy ion collisions. We interpret the 
analysis results  and argue that  color
deconfined state has been formed at highest CERN--SPS energies 
and at BNL--RHIC.}

\vskip -11cm \ \hfill CERN-TH/2001-346 \vskip 10.5cm
\section{Quark-Gluon Plasma Phase Boundary}
The signature of quark-gluon plasma is hard to identify, since the
final state observed in the laboratory always 
consists of the same particles, irrespective of the
transitional presence of the deconfined state. What changes is 
 the detailed composition of the observed 
produced particle abundance.
We consider here strangeness \cite{Raf82b}, and the pattern
of strange hadron production at SPS and RHIC
which offer, as we shall see, a compelling evidence for  quark-gluon plasma 
formation.

An expanding fireball of quark and gluons breaks up into final state
hadrons in conditions which differ from the equilibrium 
transformation explored in lattice QCD \cite{Kar00}. To understand
this behavior we need to understand the pressure of quark-gluon plasma:
\beqarl{ZQGPL1}
P_{\mathrm{QGP}}\!+\!{\cal B}=\frac{8 c_1}{45\pi^2}(\pi T)^4\! +\!
\frac{n_{\mathrm f}}{15\pi^2}
\left[\frac{7}{4}c_2(\pi T)^4\!+\!\frac{15}{2}c_3\left(
\mu_{\mathrm q}^2(\pi T)^2\! +\! \frac{1}{2}\mu_{\mathrm q}^4
\right)\right].
\eeqar
We have inserted here the appropriate quark (with $n_{\rm f}=2$) 
and gluon degeneracy.
The interactions between
quarks and gluons manifest their presence aside of the 
vacuum structure effect ${\cal B}$,
in the three coefficients $c_i\ne 1$, see \cite{Chi78}:
\beql{ICZQGP1}
c_1= 1-\frac{15\alpha_s}{4\pi}+ \cdots\,,\quad
c_2= 1-\frac{50\alpha_s}{21\pi}+ \cdots\,,\quad
c_3=1-\frac{2\alpha_s}{\pi}+ \cdots\,.
\eeq
While higher order terms in
the perturbative expansion have been obtained,
they suggest lack of convergence of the expansion
scheme formulated today.
To reproduce the lattice results, near to $T=T_{\rm c}$, the bag constant, 
${\cal B}=0.19\,{\mbox{GeV}}/{\mbox{fm}}^3$\,,
is used.

The condition $P=P_{\mbox{\scriptsize p}}-{\cal B}\to 0$, including the effect of 
motion  reads \cite{Raf00},
\beql{BPeqvS}
{\cal B}=P_{\mbox{\scriptsize p}}+
      (P_{\mbox{\scriptsize p}}+\epsilon_{\mbox{\scriptsize p}})
\frac{\kappa  v_{\mbox{\scriptsize c}}^2}{1-v_{\mbox{\scriptsize c}}^{2}}\,,\qquad 
\kappa=\frac{(\vec v_{\mbox{\scriptsize c}}\cdot \vec n)^2}{v_{\mbox{\scriptsize c}}^2}\,,
\eeq
where subscript p refers to particle pressure and energy, and
where we introduced the geometric factor $\kappa$ which
characterizes the angular relation between the surface normal vector and flow direction. 
The motion of quarks and gluons exercises additional pressure and thus 
the QGP phase remains the matter phase even below the 
static equilibrium  transition temperature.

A fireball surface region  which reaches condition \req{BPeqvS}
and continues to flow outwards
must   be torn apart. This is a collective instability and the 
ensuing disintegration of the fireball matter should be  very rapid. 
A rapidly evolving fireball which supercools is in general
 highly unstable, and we expect that
a sudden transformation (hadronization)
into confined matter can ensue in such a condition.
The situation we described could only arise  since the vacuum pressure 
term is not subject to flow and always keeps the same value.
An exploding QGP  fireball only contained by the vacuum 
supercools to $T\simeq 145$ MeV. Hadrons are produced well below
the equilibrium transition temperature $T\simeq$ 160--170 MeV.

\section{Statistical Hadronization}
\lsec{SudHad}
An extra reaction step 
`hadronization' is required to connect the properties 
of the deconfined quark--gluon matter  fireball,
and the experimental apparatus.
In this process, the quark and gluon content of the 
fireball is transferred into ultimately free
flowing hadronic particles. In hadronization,  
gluons fragment into quarks, and quarks coalesce into
hadrons. Color `freezes', 
\qgp\ excess entropy has to find a way to get away, so any additional
production is hindered. It is far from obvious that  hadron 
phase space (so called `hadronic gas')
be used consistently to describe the physics of thermal hadronization, and
we establish now the consistency criterion assuming that 
entropy production is small, or even null, in 
hadronization of entropy rich thermal QGP into entropy poor hadron
phases space.  

Using the Gibbs--Duham relation for a unit volume, 
$\epsilon+P=T\sigma+\mu \nu$\,,
the instability condition of dynamical expansion, 
 \req{BPeqvS}, takes the form:
\beql{BPeqvS1} 
-P|_{\rm h}\!=(P|_{\rm h}\!+\epsilon|_{\rm h}) {\kappa v_{\rm c}^2\over 1\!-\!v_{\rm c}^2},
\qquad
\left.{\epsilon\over \sigma}\right|_{\rm h}\!=
\left(T|_{\rm h}\!+\left.{\mu_{\rm b}\nu_{\rm b}\over \sigma }\right|_{\rm h}\right)
\left(1+{\kappa v_{\rm c}^2\over 1\!-\!v_{\rm c}^2}\right).
\eeq
Using extensive variables we obtain \bcite{Raf00}:
\beql{EBSfinal}
\left.{E\over S}\right|_{\rm h}\!=
(T|_{\rm h}+\delta T|_{\rm h})\left(1+{\kappa v_{\rm c}^2\over 1-v_{\rm c}^2}\right)\,,\quad
\delta T=\mu_{\rm b}\frac{\nu_{\rm b}}{\sigma}
=\frac{\mu_{\rm b}}{S/b}\,.
\eeq
For RHIC,  we have 
$ \delta T|_{\rm h} < 0.4$ MeV,
considering that $\mu_{\rm b}<40$ MeV and $S/b>100$; at top SPS  energy,
we have $\mu_{\rm b}\simeq 200$--250 MeV and $S/b\simeq 25$--45, and thus, 
 $\delta T|_{\rm h} \simeq 5$ MeV. 

The  usefulness of \req{EBSfinal} comes from the observation
that it implies:
\beql{EBSineq}
\left.\!{E\over S}\right|_{\rm h} > T|_{\rm h}\,.
\eeq
This is a near equality since
the geometric emissivity factor $\kappa$ is
positive and small, especially so at RHIC, and  $v_{\rm c}^2<1/3$.

The  Gibbs--Duham relation also implies:
${E/ S}+{PV/ S}=T+\delta T>T$\,.
One would think that the $PV$ term is small, since
the pressure is small as the lattice calculations suggest,
$\epsilon/P\to 7$. However,  this in principle
can be compensated by large volume of hadronization. Since the volume 
is a directly measured quantity, this hadronization 
requirement can not remain unnoticed for long. 
In fact, the HBT results today
place  a very severe  constraint on the emitter size  of the pion 
source. With realistic volume size, the $PV$  term is negligible. 
In this case, we obtain just as in \req{EBSineq}, ${E/S} \simeq T$\,.

In \req{EBSineq}, we have thus a necessary statistical hadronization constraint.
In our experience, this is practically impossible to accomplish
 in chemical equilibrium
statistical  hadronization models. The reason is  that, in 
such an approach, a rather high 
temperature $T\simeq 175$ MeV is required to accommodate the high
intrinsic entropy content of 
hadron source, but this does not drive up sufficiently the 
energy content, since for pions in equilibrium $E/S<T$.
What happens in such an approach is that a very large
hadronization volume is required, in disagreement with 
the HBT results. We conclude that the
combination of HBT result with fundamental properties of 
statistical physics  requires chemical
non-equilibrium statistical hadronization.

\section{Hadron Phase Space}
\lsec{Phasespace}
We  consider a generalization of Fermi's statistical model of 
hadron production~\bcite{Fer50,Fer53}, and consider the yield
of hadrons to be solely dictated by the study of the 
magnitude of the phase space available.

The relative number of  final state hadronic particles
freezing out from, {\eg}, a thermal quark--gluon source, is obtained
noting that the  fugacity $f_i$ of the $i$-th emitted  composite  
hadronic particle containing $k$-components is derived 
from fugacities $\lambda_k$ and phase space occupancies $\gamma_k$:
\beql{abund}
N_i\propto e^{-E_i/T_{\rm f}}f_i=e^{-E_i/T_{\rm f}}\prod_{k\in i}\gamma_k\lambda_k.
\end{equation}
In most cases, we study chemical properties of 
light quarks $\rm u,d$ jointly, though on occasion, we 
will introduce the isospin asymmetry.
As seen in \req{abund}, we study particle 
production in terms of five statistical parameters
$T, \lambda_{\rm q}, \lambda_{\rm s}, \gamma_{\rm q},  \gamma_{\rm s}$. In addition,
to describe the shape of spectra, one needs  matter flow velocity
parameters, these become irrelevant when only total particle abundances
are studied, obtained integrating  all of phase space, or equivalently
in presence of strong longitudinal flow, when we are looking at a
yield per unit of rapidity. 

The difference between the two  types of chemical parameters,
$\lambda_i$ and $\gamma_i$, is that the phase space
occupancy  factor $\gamma_{i}$ regulates the number of pairs of flavor `$i$', 
and hence applies in the same manner to particles and antiparticles, while
fugacity $\lambda_i$ applies only to particles, while $\lambda_i^{-1}$ is
the antiparticle fugacity.

The resulting yields of final state hadronic particles are most 
conveniently characterized taking the Laplace transform of the 
accessible phase space. This approach generates a  function which,
in its mathematical properties, is identical to 
the partition function:
\beql{4abis}
{\cal L}\left[e^{-E_i/T_{\rm f}}\prod_{k\in i}\gamma_k\lambda_k\right] 
 \propto \ln{\cal Z}^{\rm HG}\,.
\eeq
\req{4abis} does not require formation of a phase comprising a
gas of hadrons,  but is not inconsistent with such a step in evolution 
of the matter;  it describes  not a partition function, but just 
a look-alike object arising from the  Laplace transform of the accessible 
phase space. The final particle abundances, measured in an experiment, 
are obtained after all unstable hadronic resonances `$j$'
are allowed to disintegrate, contributing to the yields of
stable hadrons.

 The unnormalized particle multiplicities   are obtained 
differentiating \req{4abis} with respect to particle 
fugacity. The relative particle yields are simply 
given by ratios of corresponding chemical factors, weighted with
the size of the momentum phase space accepted by the experiment.
The ratios of strange antibaryons to strange baryons {\it
of same particle type\/} are, in our approach, simple functions of the quark 
fugacities.

\section{Strange Hadrons at SPS}
\lsec{SPShad}
We expect, in sudden hadronization, 
chemical non-equilibrium at hadron freeze-out.
Full chemical non-equilibrium description of 
particle yields is required to arrive at a statistically 
significant description of the data with small  error: 
\beql{chi2}
\chi^2\equiv\frac{\sum_j({R_{\rm th}^j-R_{\rm exp}^j})^2}{
({{\Delta R _{\rm exp}^j}})^2}\,.
\end{equation}
It is common to normalize the total error $\chi^2$ by the 
difference between the number of data points and parameters used,
the so called `dof' (degrees of freedom) quantity.
For systems we study, with 
a few degrees of freedom (typically 5--15),  a  statistically 
significant fit requires that $\chi^2/$dof $<1$\,.
For just a few `dof', the error should 
be as small as $\chi^2/$dof$<0.5$.
The usual  requirement  $\chi^2\to 1$ is  only applying
for very large `dof'.

\begin{table}[tb]
\captt{resultpb2}{Pb--Pb 158\agev particle ratios}{
\small
WA97 (top) and NA49 (bottom)  Pb--Pb 158\agev collision hadron ratios
compared with phase space fits.
\index{particle ratios!Pb--Pb at 158\agev}}
\vspace{-0.1cm}
\begin{center}
\begin{tabular}{lclll}
\hline\hline
\baselineskip 0.9cm
 Ratios\phantom{$\Big($}      & Ref. &  Exp. Data       & Pb$|^{\rm s,\gamma_{\rm q}}$ & Pb$|^{\rm \gamma_{\rm q}}$ \\
\hline
${\Xi}/{\Lambda}$\phantom{$\Big($}      &  \protect \cite{Kra98} &0.099 $\pm$ 0.008                     & 0.096 & 0.095\\
${\overline{\Xi}}/{\bar\Lambda}$ &  \protect \cite{Kra98} &0.203 $\pm$ 0.024      & 0.197 & 0.199\\
${\bar\Lambda}/{\Lambda}$  &  \protect \cite{Kra98} &0.124 $\pm$ 0.013            & 0.123 & 0.122\\
${\overline{\Xi}}/{\Xi}$  &  \protect \cite{Kra98} &0.255 $\pm$ 0.025             & 0.251 & 0.255\\
\hline
${\rm K^+}/{\rm K^-}$\phantom{$\Big($}              &  \protect \cite{Bor97}         &  1.80$\pm$ 0.10         & 1.746  & 1.771 \\
${\rm K}^-/\pi^-$         &       \protect \cite{Sik99}&  0.082$\pm$0.012        & 0.082  & 0.080\\
${\rm K}^0_{\rm s}/b$       &  \protect \cite{Jon96}   & 0.183 $\pm$ 0.027       & 0.192  & 0.195\\
${h^-}/b$                 &  \protect \cite{App99}     & 1.97 $\pm $ 0.1\ \      & 1.786  & 1.818 \\
$\phi/{\rm K}^-$   &  \protect \cite{Afa00}  & 0.145 $\pm$ 0.024\ \                 & 0.164  & 0.163 \\
${\bar\Lambda}/{\rm \bar p}$     &       $y=0$             &                       & 0.565  & 0.568 \\
${\rm \bar p}/\pi^-$             &       all $y$           &                       & 0.017  & 0.016 \\
\hline
 & $\chi^2$     &     & 1.6 & 1.15 \\
 &  $ N;p;r$     &    & 9;4;1& 9;5;1\\
\hline\hline
\vspace*{-0.9cm}
\end{tabular}
\end{center}
\end{table}
Turning to the Pb--Pb system at 158$A$ GeV collision energy,
we consider  particle listed  in \rt{resultpb2}, top section from the
experiment WA97, for $p_\bot>0.7$\,GeV, within a narrow
$\Delta y=0.5$ central rapidity window. Further below 
are shown  results from the large  acceptance experiment NA49, 
extrapolated by the collaboration to full $4\pi$ phase space coverage. 
The total error $\chi^2$
for the  two result columns is shown at the bottom of this table
along with the number of data points `$N$', parameters `$p$' 
used, and number of (algebraic) redundancies `$r$' connecting the 
experimental results. For $r\ne 0$, it is more appropriate 
to quote the total  $\chi^2$, since the 
statistical relevance condition is more difficult to 
establish given the constraints, but since
 $\chi^2/(N-p-r)<0.5$, we are certain to have a valid description of
hadron multiplicities. 

In second last  column, the  superscript `s' 
means that $\lambda_{\rm s}$ is fixed by strangeness balance and,  superscript 
`$\gamma_{\rm q}$', in two last columns,  means that 
$\gamma_{\rm q}=\gamma_{\rm q}^{\rm c}=e^{m_\pi/2T_{\rm f}}$, 
is fixed to maximize the entropy content in the hadronic phase space.
The fits presented  are obtained with latest NA49 experimental results, \ie,
have updated  ${\rm h}^-/b$, newly published $\phi$ yield \bcite{App99}, and we predict
the ${\bar\Lambda}/{\rm \bar p}$ ratio.  $b$ is here the number of baryon
participants, and  $\rm h^-=\pi^-+K^-+\bar p$ is the yield of stable negative
hadrons which includes pions, kaons and antiprotons.
We see, comparing the two columns, that strangeness conservation 
(enforced in second last column) 
is consistent with the experimental data shown,  enforcing it 
does not change much the results for particle multiplicities. 

The six  parameters ($T, v_{\rm c}, \lambda_{\rm q}, 
\lambda_{\rm s}, \gamma_{\rm q}, \gamma_{\rm s}$)
describing the  particle abundances
are shown in the top section of \rt{fitqpbs}.
Since the results of the WA97 experiment  are not covering the full phase space, 
there is a reasonably precise value found  for  one velocity parameter,
 taken to be the spherical surface flow velocity $v_{\rm c}$ of the
fireball hadron source.

A value $\lambda_{\rm s}^{\rm Pb}\simeq 1.1$
characteristic for a source of freely movable  strange quarks with
balancing strangeness
in presence of strong Coulomb potential 
 \cite{Let99d}, {\ie}, with $\tilde\lambda_{\rm s}=1$, is obtained. 
Since all chemical non equilibrium studies of the Pb--Pb 
system converge to the case of maximum entropy, 
we have  presented the results with fixed 
$\gamma_{\rm q}=\gamma_{\rm q}^{\rm c}=e^{m_\pi/2T_{\rm f}}$. 
The large values of $\gamma_{\rm q}>1$ 
confirm the need to hadronize the excess entropy of the 
QGP possibly formed. This value is derived from both 
 the specific  negative hadron  ${\rm h}^-/b$ abundance  
and from the relative strange hadron yields.

\begin{table}[tb]
\captt{fitqpbs}{statistical model parameters}{
\small
Upper section: statistical model parameters
which best describe the experimental results for
Pb--Pb data seen in Fig.\,\ref{Tab:resultpb2}.
Bottom section: energy per entropy, antistrangeness, net strangeness
 of  the full hadron phase space characterized by these
statistical parameters. In column two, we fix $\lambda_{\rm s}$ by requirement of 
strangeness conservation, and in this and next column we fix 
$\gamma_{\rm q}=\gamma_{\rm q}^{\rm c}$.
Superscript $^*$ indicates values which are result of a constraint.
\index{chemical parameters!Pb--Pb system}
}
\vspace{0.cm}\begin{center}
\begin{tabular}{lcc}
\hline\hline
\phantom{$\Big($}                           & Pb$|_v^{\rm s,\gamma_{\rm q}}$ & Pb$|_v^{\rm \gamma_{\rm q}}$\\
\hline
$T$ [MeV]             &  151 $\pm$ 3      &  147.7 $\pm$ 5.6           \\
$v_{\rm c}$           & 0.55 $\pm$ 0.05   & 0.52 $\pm$ 0.29        \\
$\lambda_{\rm q}$     & 1.617 $\pm$ 0.028 & 1.624 $\pm$ 0.029       \\
$\lambda_{\rm s}$     & 1.10$^*$         & 1.094 $\pm$ 0.02         \\
$\gamma_{\rm q}$  & ${\gamma_{\rm q}^{\rm c}}^*=e^{m_\pi/2T_{\rm f}}$=1.6  &${\gamma_{\rm q}^{\rm c}}^*=e^{m_\pi/2T_{\rm f}}$=1.6\\
$\gamma_{\rm s}/\gamma_{\rm q}$   & 1.00 $\pm$ 0.06  & 1.00 $\pm$ 0.06         \\
\hline 
$E/b$[{\small GeV}]    &  4.0    &     4.1      \\
${s}/b$               & 0.70 $\pm$ 0.05  & 0.71 $\pm$ 0.05         \\
$E/S$[{\small MeV}]   & 163 $\pm$ 1    & 160 $\pm$ 1           \\
$({\bar s}-s)/b\ \ $  & 0$^*$            &  0.04 $\pm$ 0.05    \\ 
\hline\hline
\vspace*{-0.9cm}
\end{tabular}
\end{center}
\end{table}

The fits shown 
satisfy comfortably the statistical hadronization constraint 
that $E/S>T$ discussed in \res{SudHad}.  We see also that near strangeness 
balance  is obtained as result of the fit. 

One of the interesting  quantitative results of this analysis 
is shown in the bottom section of \rt{fitqpbs}: the
 yield of strangeness per baryon, $s/b\simeq 0.7$\,. 
As we will show in \rf{PLSBLAMQ}, the expected equilibrium yield is
 $s/b|^{\rm QGP}\simeq 1.4$\,, and thus we have $\gamma_{\rm s}^{\rm QGP}\simeq 0.5$
at SPS.  Since the occupancy values in \rt{fitqpbs} are derived
from hadron phase space, we thus find  
$\gamma_{\rm s}^{\rm HG}/\gamma_{\rm s}^{\rm QGP}\simeq 3$\,.

\section{Strangeness at RHIC}
\lssec{SRHIC}
In the likely event that the QGP formed at RHIC evolves
towards strangeness chemical 
equilibrium abundance, or possibly even exceeds it, we should 
expect a noticeable over occupancy  of strangeness as measured in terms 
of chemical equilibrium  final state hadron abundance. Because much of the
strangeness is in the baryonic degrees of freedom, the  kaon to pion ratio 
should appear suppressed,  compared to SPS results.
An even more penetrating effect of the hadronization of strangeness rich 
QGP at RHIC is the abundant formation of strange 
baryons and antibaryons \bcite{Raf99a}. This high phase space occupancy is
one of the requirements for  the enhancement of multistrange (anti)baryon
production, which is an important hadronic signal of
QGP phenomena \bcite{Raf82a}. In particular, we
hope that hadrons produced in phase space with a small probability, 
such as $\Omega,\,\overline\Omega$, will be observed with a 
yield above statistical hadronization expectations, 
continuing the trend seen at SPS.

The RHIC data we consider are were obtained in $\sqrt{s_{\rm NN}}=130$ GeV run
at the central rapidity region where, due to approximate  
longitudinal scaling, the effects of flow cancel and 
we can evaluate the full phase space yields
in order to obtain particle ratios. We do not fit trivial results
such as $\pi^+/\pi^-=1$, since the large hadron yield combined with the
flow of baryon isospin asymmetry towards the fragmentation rapidity region 
assures us that this result will occur to a great precision.
We also do not use the results for
$\rm K^*, \bar K^*$ since these yields depend on the degree of rescattering of
resonance decay products. 
The data we consider has been reported
by the STAR collaboration of Summer 2001, and where available 
is combined with data of PHENIX, BRAHMS, PHOBOS, for more
discussion of the data origin, see \bcite{Bra01}.
We assume, in our fit in \rt{RHIChad}, that the multistrange
weak interaction cascading $\Xi\to \Lambda$, in the STAR
result we consider,  is  cut by  vertex discrimination
and thus we use these yields without a further correction.

We first consider what  experimental hadron yield results shown 
in \rt{RHIChad} imply about 
total strangeness yield in the RHIC-130 fireball. 
We begin with the yield of strange quarks contained in hyperons. 
We have, in singly strange hyperons, 1.5 times
the yield observed in $\Lambda$,
since $\Sigma^{\pm}$ remain unobserved. Also, accounting for the doubly strange
$\Xi^-$ which are half of the all $\Xi$, and contain two strange quarks, we have:
\[
{\langle s\rangle_{\rm Y}\over\rm  h^-}=1.5\cdot 0.059+2\cdot 2\cdot  0.195\cdot 0.059=0.133\,.
\]
Allowing for the unobserved $\Omega$ at the theoretical rate,
this number increases to $\langle s\rangle_{\rm Y}/\rm h^- =0.14$. Repeating the same 
argument for antihyperons the result is $0.10$. $\rm s$ and $\rm \bar s$ content in kaons is 
four times that in $\rm K_S$ and thus the total strangeness yield is
\[
{\langle\rm  s + \bar s\rangle\over\rm  h^-}=0.76\,,
\]
with 32\% of this yield contained in hyperons and antihyperons. 
This amounts to about 1.5 fold enhancement compared to highest SPS 
energies, with (multi)strange hyperons and antihyperons being a very 
important component.

\begin{table}[t]
\captt{RHIChad}{RHIC 130 analysis}{ 
\small
Fits of central rapidity hadron ratios for RHIC  $\sqrt{s_{\rm NN}}=130$ GeV run.
Top section: experimental results, followed by chemical parameters, 
physical property of the phase space, and the fit error. Columns: data, full non-equilibrium
fit, nonequilibrium fit constrained by strangeness conservation and supersaturation 
of pion phase space,  and in the last column, equilibrium fit constrained by 
strangeness conservation, upper index $^*$ indicates quantities fixed by these 
considerations.
\index{particle ratios!Pb--Pb at $\sqrt{s_{\rm NN}}=130$ GeV }
\index{chemical parameters!Pb--Pb at $\sqrt{s_{\rm NN}}=130$ GeV}}
\vspace*{0.cm}
\begin{center}
\begin{tabular}{lcccc}
\hline\hline
                                        & Data  & Fit& Fit         & Fit$^{\rm eq}$   \\
                                        &       &    & $\rm s-\bar s=0$&  $\rm s-\bar s=0$\\
\hline
$ \rm{\bar p}/p$                            &0.64\ $\pm$0.07\ & 0.637 & 0.640 &  0.587 \\
$\rm{\bar p}/h^-$                          &                & 0.068 & 0.068 &  0.075 \\
${\overline\Lambda}/{\Lambda}$          &0.77\ $\pm$0.07\ & 0.719 & 0.718 &  0.679  \\
$\rm{\Lambda}/{h^-}$                       & 0.059$\pm$0.001 & 0.059 & 0.059 &  0.059  \\
$\rm{\overline\Lambda}/{h^-}$              & 0.042$\pm$0.001 & 0.042 & 0.042 &  0.040  \\
${\overline{\Xi}}/{\Xi}$                &0.83\ $\pm$0.08\ & 0.817 & 0.813 &  0.790  \\
${\Xi^-}/{\Lambda}$                     & 0.195$\pm$0.015 & 0.176 & 0.176 &  0.130  \\
${\overline{\Xi^-}}/{\overline\Lambda}$ & 0.210$\pm$0.015 & 0.200 & 0.200 &  0.152  \\
$\rm{K^-}/{K^+}$                           &0.88\ $\pm$0.05\ & 0.896 & 0.900 &  0.891  \\
$\rm{K^-}/{\pi^-}$                         & 0.149$\pm$0.020 & 0.152 & 0.152 &  0.145  \\
$\rm{K_S}/{h^-}$                           & 0.130$\pm$0.001 & 0.130 & 0.130 &  0.124  \\
${\Omega}/{\Xi^-}$                      &                & 0.222 & 0.223 &  0.208 \\
${\overline{\Omega}}/{\overline{\Xi^-}}$&                & 0.257 & 0.256 &  0.247 \\
${\overline{\Omega}}/{\Omega}$          &                & 0.943 & 0.934 &  0.935    \\
\hline
$T$                                       &           &158$\pm$ 1   & 158$\pm$ 1     &  183$\pm$ 1      \\
$\gamma_{\rm q}$                            &      &1.55$\pm$0.01 & 1.58$\pm$0.08  &  1$^*$    \\
$\lambda_{\rm q}$                           &      &1.082$\pm$0.010& 1.081$\pm$0.006& 1.097$\pm$0.006      \\
$\gamma_{\rm s}$                            &       & 2.09$\pm$0.03 &  2.1$\pm$0.1   &  1$^*$    \\
$\lambda_{\rm s}$                           &     &$\!$1.0097$\pm$0.015$\!$&  1.0114$^*$     & 1.011$^*$ \\
\hline
$E/b$[{\small GeV}]\ \                        &   &  24.6& 24.7  &  21    \\
$s/b$                                         &   &  6.1 &  6.2  &  4.2  \\
$S/b$                                         &   & 151  &  152  &  131  \\
$E/S$[{\small MeV}]\ \                        &   & 163  &  163  &  159  \\
\hline
$\chi^2/$dof                                 &   & 2.95/($10\!-\!5$)  &2.96$\!$/$\!$($10\!-\!4$)  & 73/($10\!-\!2$)  \\
\hline\hline
\vspace*{-0.9cm}
\end{tabular}
\end{center}
\end{table}

We now discuss the fits of RHIC-130 results.
In the last column in  \rt{RHIChad},
the chemical equilibrium fit, the large  $\chi^2$
originates in the inability to account for multistrange 
$\overline\Xi,\,\Xi$. Similar results are 
presented in Ref. \bcite{Bra01}, 
which work does not include multistrange hadrons.
This equilibrium fit yields $E/S=159\,\mbox{MeV}<T=183$ MeV
contradicting the conditions we discussed in depth in \res{SudHad}.
On the other hand, the chemical nonequilibrium fits come out
to be in near perfect agreement with data, and consistent with the QGP 
statistical hadronization
picture:  $E/S=163>T=158$ MeV and $\gamma_{\rm s},\gamma_{\rm q}>1$\,.
The value of the hadronization temperature $T=158$ MeV is 
 below the  central expected equilibrium phase transition
temperature, and this  hadronization  temperatures at RHIC is  
consistent with sudden breakup of a supercooled QGP fireball.
The inclusion of the  
yields of multistrange antibaryons in the RHIC data analysis, 
along with allowance for chemical non-equilibrium ($\gamma\ne 1$), allows to
discriminate the different (chemical equilibrium/nonequilibrium) reaction scenarios. 

The  value of the thermal energy content $E/b=25$ GeV, seen in \rt{RHIChad}, is in very good 
agreement with expectations once we allow for the kinetic energy content
associated with longitudinal and transverse motion. The energy of each particle 
is `boosted' with the factor $\gamma_\bot^v\cosh y_\parallel$. 
For $v_\bot=c/\sqrt{3}$, we have $\gamma_\bot^v=1.22$.
The longitudinal flow range is about $\pm 2.3$ rapidity units, 
according to PHOBOS results.
To obtain the energy increase due to longitudinal flow, we have to multiply 
by the average,  
$\int dy_\parallel \cosh y_\parallel/y_\parallel\to\sinh(2.3)/2.3=2.15$, 
for a total average increase in energy by 
factor 2.62, which takes the full energy content to 
$E^v/b\simeq 65$ GeV as expected. This consistency reassures that
we have a physically relevant fit of the data, which respects 
overall energy conservation.

We analyze next the strangeness content, $s/b=6$, seen  \rt{RHIChad}.  The fully 
equilibrated QGP phase space would have yielded  8.6 strange quark pairs per baryon
at $\lambda_{\rm q}=1.08$,  as is shown  in \rf{PLSBLAMQ} below. Thus
$\gamma_{\rm s}^{\rm QGP}=6/8.6\simeq 0.7$\,, which is greater than the value 
0.5 we found for  the SPS energy range. Using the
fitted value  $\gamma_{\rm s}^{\rm HG}=2.1$, we find  again, just like we
found in case of SPS, 
$\gamma_{\rm s}^{\rm HG}/\gamma_{\rm s}^{\rm QGP}\simeq 3$.  The fact that the 
strangeness phase space in QGP is not fully saturated is,
on a second careful look, in qualitative agreement
with  kinetic strangeness theory predictions  \bcite{Raf99a}, 
adjusting  our study to the observed RHIC-130 run
conditions.

\section{Strangeness as Signature of Deconfinement}
\lsec{SsigQGP}
We consider the ratio of equilibrium strangeness density, arising in the
Boltzmann gas limit,  to the baryon density 
in a QGP fireball. To first approximation,
perturbative thermal QCD corrections cancel in the ratio.
For $m_{\rm s}=200$ MeV and $T=150$ MeV, we have:
\beql{sdivb1}
{s\over b}\simeq  
  \gamma_{\rm s}^{\mathrm{QGP}}\frac{0.7}
          {\ln \lambda_{\rm q} +{{(\ln \lambda_{\rm q})^3}/{\pi^2}}}\,.
\eeq

The relative yield $s/b$ is mainly dependent on the value of $\lambda_{\rm q}$,
with only a slow temperature dependence contained in the coefficient $0.7$ 
in \req{sdivb1}. The
light quark fugacity  $\lambda_{\rm q}$
is usually independent of the strategy of data analysis, as is also seen in
\rt{RHIChad}. 

\myfig{0.3}{10}{PLSBLAMQP}{0}{-3.7}{-0.9}{tb}
\capt{PLSBLAMQ}{S/b yield}{
\small
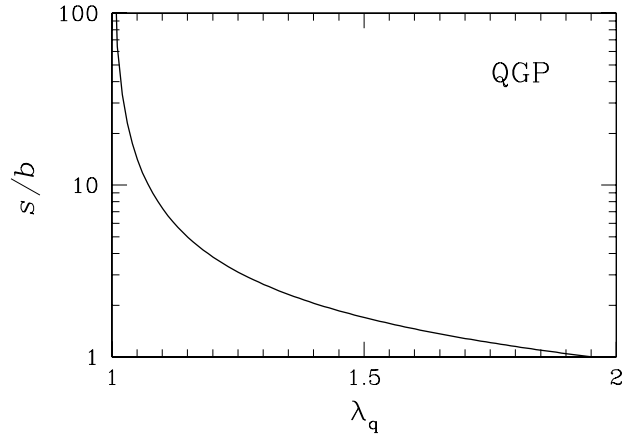
Strangeness yield per baryon as function of $\lambda_{\rm q}$ 
in equilibrated quark-gluon plasma.
}
\end{figure}

At  top SPS energy where $\lambda_{\rm q}\simeq 1.6$, we see in \rf{PLSBLAMQ} that 
the equilibrium strangeness yield is at 1.5 strange pairs per participating baryon.
The actual experimental yield, 0.7, see \rt{fitqpbs},
is half as large as  in an equilibrated QGP, 
but is 2.5 times the  yield in p--p reactions. 
At the RHIC 130 GeV run, the value $\lambda_{\rm q}=1.08$, see \rt{RHIChad}, 
and the specific strangeness yield in a
QGP fireball at equilibrium is  an order of magnitude greater than 
currently observed at SPS top energy. As discussed, the actual yield corresponds 
to 70\% of the full phase space. A further increase towards equilibrium 
yield at higher RHIC-200 energy range is expected. 
The remarkable feature of the RHIC situation is that much of  strangeness 
enhancement is  found in the (multistrange) baryon abundance. 
Given the large 
strangeness per baryon ratio, \rf{PLSBLAMQ},  baryons and antibaryons produced 
from QGP at
RHIC have been predicted to be mostly  strange \bcite{Raf99a}.

The high specific strangeness yield $s/b$ is a clear indicator for
the extreme conditions reached in heavy ion collisions. An equally
interesting observable is the occupancy of the  strangeness phase space.
In sudden hadronization,  
$V^{\mathrm{HG}}/ V^{\mathrm{QGP}}\simeq 1$, 
the growth of volume is negligible,  
$T^{\mathrm{QGP}}\simeq T^{\mathrm{HG}}$, the temperature 
is maintained across the hadronization front, and the chemical occupancy 
factors in both states of matter accommodate the different magnitude of the particle
phase space. In this case, the QGP  strangeness when `squeezed' into the
smaller HG phase space results in 
${\gamma_{\rm s}^{\mathrm{HG}}/\gamma_{\rm s}^{\mathrm{QGP}}}\simeq 3$\,,
which is of the same magnitude as the unfrozen color degeneracy. 
This  theoretical expectation is indeed  observed both at top SPS and RHIC-130 run,
as we have reported. 

For the top SPS energy range this interesting result could be ignored, since
accidentally  $\gamma_{\rm s}^{\mathrm{HG}}/ \gamma_{\rm q}^{\mathrm{HG}}\simeq 1$. Thus
one can also model the hadronization at SPS energy in terms of an
equilibrium hadronization model. The pion enhancement associated with 
the high entropy phase can be accommodated by use of two temperatures, one
for the determination of absolute particle yields, and another for 
determination of the spectral shape. Such an approach has
 similar number of parameters, and comparable predictive power, the only
inconsistency (with HBT) is the large volume required, as we have discussed.
However, the  condition, 
$\gamma_{\rm s}^{\mathrm{HG}}/ \gamma_{\rm q}^{\mathrm{HG}}\simeq 1$,
is  not present at the  RHIC energy range, where already at RHIC-130 the
hadron phase space occupancy for strangeness is significantly larger than
for light quarks, see  \rt{RHIChad}.  It is the inclusion of the  
yields of multistrange antibaryons in the RHIC data analysis, which
leads to this result.

We see, at SPS and at RHIC, 
considerable convergence of the hadron production 
around properties of suddenly hadronizing entropy and
strangeness rich  QGP. The QGP phase strangeness occupancy rises from 
$\gamma_{\rm s}^{\rm QGP}\simeq 0,5$ at SPS $\sqrt{s_{\rm NN}}=17.2$ GeV, to  
$\gamma_{\rm s}^{\rm QGP}\simeq 0.7$ at $\sqrt{s_{\rm NN}}=130$ GeV, and
there is still space for a further strangeness yield rise at highest RHIC 
energy $\sqrt{s_{\rm NN}}=200$ GeV.
In conclusion: the deconfined thermal QGP phase manifests itself through  
its gluon content,  which generates in thermal collision processes a 
clear strangeness  fingerprint of QGP.

\subsection*{Acknowledgments}
Work supported in part by a grant from the U.S. Department of
Energy,  DE-FG03-95ER40937. Laboratoire de Physique Th\'eorique 
et Hautes Energies, University Paris 6 and 7, is supported 
by CNRS as Unit\'e Mixte de Recherche, UMR7589.




\begin{thebibliography}{99}
\small
\bibitem{Raf82b}
J.~Rafelski and B.~{M\"uller}, 1982.
 Strangeness production in the quark--gluon plasma.
 {\em Phys. Rev. Lett.}, {\bf 48}, 1066.
 See: {\it Phys. Rev. Lett.}, {\bf 56}, 2334E (1986).

\bibitem{Kar00}
F.~Karsch, E.~Laermann, and A.~Peikert, 2000.
 The pressure in 2, 2+1 and 3 flavour {QCD}.
 {\em Phys. Lett. {\rm B}}, {\bf 478}, 447.

\bibitem{Chi78}
S.A. Chin, 1978.
 Transition to hot quark matter in relativistic heavy-ion collision.
 {\em Phys. Lett. {\rm B}}, {\bf 78}, 552.

\bibitem{Raf00}
J.~Rafelski and J.~Letessier, 2000.
 Sudden hadronization in relativistic nuclear collisions.
 {\em Phys. Rev. Lett.}, {\bf 85}, 4695.



\bibitem{Fer50}
E.~Fermi, 1950.
 High-energy nuclear events.
 {\em Prog. Theo. Phys.}, {\bf 5}, 570.

\bibitem{Fer53}
E.~Fermi, 1953.
 Multiple production of pions in nucleon--nucleon collisions at
  cosmotron energies.
 {\em Phys. Rev.}, {\bf 92}, 452.


\bibitem{Kra98}
I.~{Kr\'alik, {\it et al.}, WA97 collaboration}, 1998.
 {$\Lambda,\ \Xi$} and {$\Omega$} production in {Pb--Pb} collisions at
  158\agev.
 {\em Nucl. Phys. {\rm A}}, {\bf 638}, 115.

\bibitem{Bor97}
C.~{Bormann, {\it et al.}, NA49 collaboration}, 1997.
 Kaon, lambda and antilambda production in {Pb+Pb} collisions at 158
  {GeV} per nucleon.
 {\em J. Phys. {\rm G} Nucl. Part. Phys.}, {\bf 23}, 1817.

\bibitem{Sik99}
F.~{Sikl\'er {\it et al.}, NA49 collaboration}, 1999.
 Hadron production in nuclear collisions from the {NA}49 experiment at
  158\agev.
 {\em Nucl. Phys. {\rm A}}, {\bf 661}, 45c.

\bibitem{Jon96}
P.G. {Jones {\it et al.}, NA49 collaboration}, 1996.
 Hadron yields and hadron spectra from the {NA49} experiment.
 {\em Nucl. Phys. {\rm A}}, {\bf 610}, 188c.

\bibitem{App99}
H.~{{Appelsh\"auser} {\it et al.}, NA49 collaboration}, 1999.
 Baryon stopping and charged particle distributions in central {Pb+Pb}
  collisions at 158 {GeV} per nucleon.
 {\em Phys. Rev. Lett.}, {\bf 82}, 2471.

\bibitem{Afa00}
S.V. {Afanasev {\it et al.}, {NA49} collaboration}, 2000.
 Production of $\phi$ mesons in p+p, p+{Pb} and central {Pb+Pb}
  collisions at {$E_{\rm beam}$} = 158\agev.
 {\em Phys. Lett. {\rm B}}, {\bf 941}, 59.


\bibitem{Let99d}
J.~Letessier and J.~Rafelski, 1999.
 Diagnostics of {QGP} with strange hadrons.
 {\em Acta Phys. Pol. {\rm B}}, {\bf 30}, 3559.

\bibitem{Raf99a}
J.~Rafelski and J.~Letessier, 1999.
 Expected production of strange baryons and antibaryons in baryon-poor
  {QGP}.
 {\em Phys. Lett. {\rm B}}, {\bf 469}, 12.

\bibitem{Raf82a}
J.~Rafelski, 1982.
 Formation and observables of the quark gluon plasma.
 {\em Phys. Rep.}, {\bf 88}, 331.

\bibitem{Bra01}
P.~Braun-Munzinger, D.~Magestro, K.~Redlich, and J.~Stachel, 2001.
 Hadron production in {Au--Au} collisions at {RHIC}.
 {\em Phys. Lett. {\rm B}}, {\bf 518}, 41.


\end{thebibliography}
\end{document}